\documentclass[11pt,twoside,a4paper]{article}
\usepackage[T1]{fontenc}
\usepackage{graphicx}
\usepackage{amsmath}
\usepackage{amssymb}
\usepackage{hyperref}
\usepackage{url}
\usepackage{siunitx}
\usepackage{rotating}
\usepackage{longtable}
\usepackage{tabularx}
\usepackage{geometry}
\usepackage{pdflscape}

\title{X-ray properties of two complementary samples of intermediate Seyfert galaxies}
\date{}
\author{Benedetta Dalla Barba $^{1,2,*}$, Luigi Foschini $^{2}$, Marco Berton $^{3}$, \\
Luca Crepaldi $^{4}$ and Amelia Vietri $^{4}$}

\begin{document}
\maketitle
{\scriptsize{\noindent $^{1}$ \quad \noindent Università degli studi dell'Insubria, 22100 Como, Italy\\
$^2$ \quad Osservatorio Astronomico di Brera, Istituto Nazionale di Astrofisica (INAF), 23807 Merate, Italy\\
$^3$ \quad European Southern Observatory (ESO), 19001 Santiago de Chile, Chile\\
$^4$ \quad Dipartimento di Fisica e Astronomia, Università di Padova, 35122 Padova, Italy\\
$^*$ \quad Correspondence: benedetta.dallabarba@inaf.it}}

\begin{abstract}
We present the X-ray spectral analysis of two complementary sets of intermediate Seyfert galaxies (ISs). Analyzing X-ray data, we estimate the hydrogen abundance $N_H$ and test its connection with the [O III] luminosity acquired from optical observations. The results confirm the conclusions drawn in a previous study concerning the lack of a direct correlation between the obscuration measure ($N_H$) and the intrinsic characteristics of the active nuclei ([O III] luminosity). Instead, we validate the existence of a correlation between the Seyfert type and the $N_H$ parameter, employing a separation threshold of approximately 10$^{22}$ atoms cm$^{-2}$. Simultaneously, our findings align with prior research, corroborating the relationship between X-ray luminosity and the [O III] luminosity.
\end{abstract}


\section{Introduction}
Intermediate Seyfert galaxies (ISs) \cite{Oster76,Oster77} represent a category within active galactic nuclei (AGN) that, according to the Unified Model (UM) \cite{Keel,Antonucci,Urry}, are viewed at intermediate angles. As~a result, they exhibit a partial level of obscuration originating from dust surrounding the nucleus. Their optical spectra display a composite profile for permitted lines (e.g., H$\alpha$ and H$\beta$), comprising both a narrow and a broad component. These components stem from the narrow-line region (NLR) and the broad-line region (BLR), respectively. The~NLR tends to be more extended, whereas the BLR is concentrated within a dusty torus. Consequently, the~spectra of Seyferts at greater inclinations, such as type 2 (Sy2), exhibit solely the narrow component, whereas those at smaller inclinations, such as type 1 (Sy1), exclusively reveal the broad component. ISs can be further classified based on either an escalating obscuration sequence or an increasing prevalence of the narrow-line component in relation to the broad one: Sy1.2, Sy1.5, Sy1.8, and~Sy1.9 (for a detailed discussion, refer to \cite{DallaBarba}).

As inclination angles decrease to their minimum, going from Sy2 to Sy1, the~relevance of the $\gamma$-ray emission component becomes more pronounced. Among~the various types of AGN, one---blazars---exhibits prominent $\gamma$-ray emissions. In~this case, the~relativistic jets are aligned, or~almost aligned, with~the observer's line of sight, intensifying the emission due to the beaming effect. Conversely, for~the other AGN categories, as~for the Seyferts, the~presence of the $\gamma$-ray component is expected to be absent or nearly absent.
However, there have been recent reclassifications, such as J$2118-0732$, identified as the first non-local IS capable of maintaining relativistic jets detectable at $\gamma$-ray energies~\cite{Jarvela20}. This atypical behavior could be explained by introducing the concept of misaligned jets. Approximately 2.8\% of $\gamma$-ray-detected AGN exhibit these misaligned jets~\cite{Foschini22}. Consequently, the~scarcity of $\gamma$-ray-emitting ISs might be attributable to instrumental limitations linked to the faintness of high-energy emission or to intrinsic processes altering the line profile, which are unrelated to~obscuration.

To explore these contrasting scenarios, our focus centers on two complementary sets of ISs, chosen based on the selection criteria detailed in \cite{DallaBarba}. We categorize these groups as ``jetted'' and ``non-jetted'' objects. Our aim is to find some spectral quantities that separate between the two classes, independently from the obscuration level proposed by the UM. In \cite{DallaBarba}, we concentrated on the optical spectra of these sources, particularly focusing on [O III]$\lambda$5007, H$\alpha$, and~H$\beta$. The~oxygen line serves as an indicator of the intrinsic properties of the AGN, in~particular as a tool for the disk luminosity~\cite{Heckmann,Risaliti11,Marin}, even if its utility is subject to certain limitations~\cite{Hass,Baum}. The~H$\alpha$ and H$\beta$ lines enable the calculation of the Balmer decrement, offering an obscuration measure. However, the~reliability of this quantity has been extensively debated by several authors~\cite{Binette,Nagao,Schnorr}. These authors highlighted the limitations of assuming a theoretical value of 2.9 for the H$\alpha$/H$\beta$ ratio (case B recombination). In~particular, \cite{Schnorr} found a range of 2.5--6.6 for the theoretical value, concluding that a single set of extinction properties cannot be assumed for all the BLRs. As~an alternative, the \mbox{X-ray} spectrum provides a more dependable measure of obscuration through the hydrogen column density parameter ($N_H$). 

In \cite{DallaBarba}, we did not find a clear separation between the jetted and non-jetted sources in the [O III]/H$\beta$-L$_{\mathrm{[O III]}}$ plot which means that the central engine’s structure in Seyfert galaxies may be consistent across all types and the obscuration-dependent scenario remains the most probable solution. On~the other hand, in~the central panel of Figure~6 of \cite{DallaBarba}, we compared the two IS classification methods introduced by~\cite{Whittle,Netzer}, but~a clear separation between the Seyfert sub-types is not evident. This can imply two scenarios; in~the first, the jetted sources do not always correspond to Sy1.2/1.5, but~can sometimes be related to Sy1.8/1.9. In~the second, the~selection methods based on hard X-rays and $\gamma$-rays only bias the classification. In~this case, some hard-X-ray-selected objects can be Sy1.2/1.5 and $\gamma$-ray-selected AGN Sy1.8/1.9, which in turn can be explained through the misaligned jets introduced before. To~confirm or reject these results, we need a more reliable measure of the intrinsic absorption that can be obtained using X-ray spectroscopy as stated in the previous~paragraph.

The structure of this paper is as follows: Section~\ref{sec2} details the Data Reduction and Analysis, Section~\ref{sec3} presents the Results, and~Section~\ref{sec4} outlines the Summary and Conclusions. We adopt a standard $\Lambda$CDM cosmology with the Hubble constant and the cosmological parameters, respectively, of:  H$_0$ = 73.3 km s$^{-1}$ Mpc$^{-1}$, $\Omega_{matter}$ = 0.3, and~$\Omega_{vacuum}$ = 0.7~\cite{Riess}.

\section{Data Reduction and~Analysis}\label{sec2}
\subsection{Sample and Data~Reduction}\label{sec2.1}
The dataset comprises 38 sources from the \emph{Swift}-BAT AGN Spectroscopic Survey (BASS) and 11 sources from the fourth Fermi Gamma-ray Large Area Telescope (4FGL). In~the former, sources were selected using their emission in hard X-rays, suggesting a reflection of emitted radiation from the accretion flow towards the dusty region. This method tended to show a preference for higher Seyfert types (Sy2, Sy1.9, and~Sy1.8). Conversely, the~latter is $\gamma$-ray-selected with sources exhibiting a bias toward lower Seyfert types (Sy1, Sy1.2, and~Sy1.5) due to the beaming effect. All the optical spectra come from the Sloan Digital Sky Survey (SDSS); for~a comprehensive description, refer to Paper~I. 

For the X-ray counterpart, 42 sources presented {\it Swift} data, while for the remainder we employed higher-resolution observations from {\it Chandra} and {\it XMM-Newton}. Refer to Table~\ref{taba1} in the Appendix~\ref{app1} for specific details. Our objective is to measure the $N_H^{\mathrm{int}}$ parameter (intrinsic component of $N_H$), without~requiring a comprehensive understanding of the source's physics during the fitting~process. 

Data retrieval was performed from the High Energy Astrophysics Science Archive Research Center (HEASARC), followed by the utilization of the {\it Swift}-XRT pipeline version 0.13.7 and CALDB files updated on 24 July 2023 to extract the spectra. Similarly, {\it Chandra} data were processed using dedicated reduction tools like ciao-4.15 and a calibration database version 4.10.7 (14 September 2023). For~{\it XMM-Newton}, the same with SAS 21.0.0 (xmmsas\_20230412\_1735-21.0.0) for the extraction\footnote{\url{https://www.cosmos.esa.int/web/xmm-newton/sas-threads}}, and~the respective calibration datasets updated on 5 October 2023.  For~{\it Swift} and {\it XMM-Newton} data, the~source extraction radius was 40'', while the background spectra were extracted from source-free circular regions of 1' for each observation. For~{\it Chandra} observations, the extraction area was of 6'' for the sources and of 1' for the~background. 

Subsequently, for~each source we summed the available spectra to increase the statistics and binned the data into groups of 20 counts/bin to employ the $\chi^2$ statistics. This procedure was applied to all cases except J$0958+3224$ (3C 232). For~this particular source, we applied likelihood statistics for Poisson-distributed data using the {\tt c-stat} package within~Xspec.

\subsection{Data~Analysis}\label{sec2.2}
The analysis was conducted utilizing Xspec version 12.13.0c. As~mentioned earlier, our primary objective was to estimate $N_H^{\mathrm{int}}$ in units of atoms cm$^{-2}$. Initially, we employed a simple model comprising galactic absorption ({\tt tbabs}) combined with either a single redshifted power-law ({\tt zpo}) or a broken power-law ({\tt bknpo}) to represent AGN emission, denoted as {\tt tbabs*zpo} or {\tt tbabs*bknpo}. The galactic absorption data ($N_H^{\mathrm{gal}}$) were acquired through HEASARC's specialized tool\footnote{\url{https://heasarc.gsfc.nasa.gov/cgi-bin/Tools/w3nh/w3nh.pl}} \cite{Heasarc}. This initial model was suitable for Seyferts where internal obscuration plays a negligible role, such as Sy1. For~these objects, the~value reported in the plots refers to the sole external galactic absorption component ($N_H^{\mathrm{gal}}$). However, in~Sy2 and certain ISs, internal obscuration from the dust surrounding the nucleus ($N_H^{\mathrm{int}}$) could be significant. Observationally, these sources indicate a deficiency in the residuals at low energies (<1 keV) when fitted solely with galactic absorption and a power-law or broken power-law model. Consequently, we modified the model by incorporating an internal obscuration component ({\tt ztbabs}), adopting the following structure: {\tt tbabs*ztbabs*zpo} or {\tt tbabs*ztbabs*bknpo}. When this representation proved to be unsatisfactory, we resorted to a more complex model, composed of {\tt tbabs*(zpo+ztbabs*zpo)}, adapted from~\cite{Zhao}. In~this model, the~first power-law is obscured through internal processes and represents the emission from the central regions, such as the accretion disk. The~second power-law emission is dimmed only by the galactic dust and reflects the scattered emission. Furthermore, whenever fluctuations surpassing >3$\sigma$ were detected around 6.4 KeV, we incorporated a {\tt zgauss} component to reproduce the Fe K$\alpha$ line.
 
\begin{figure}[h]
\includegraphics[width=13.6 cm]{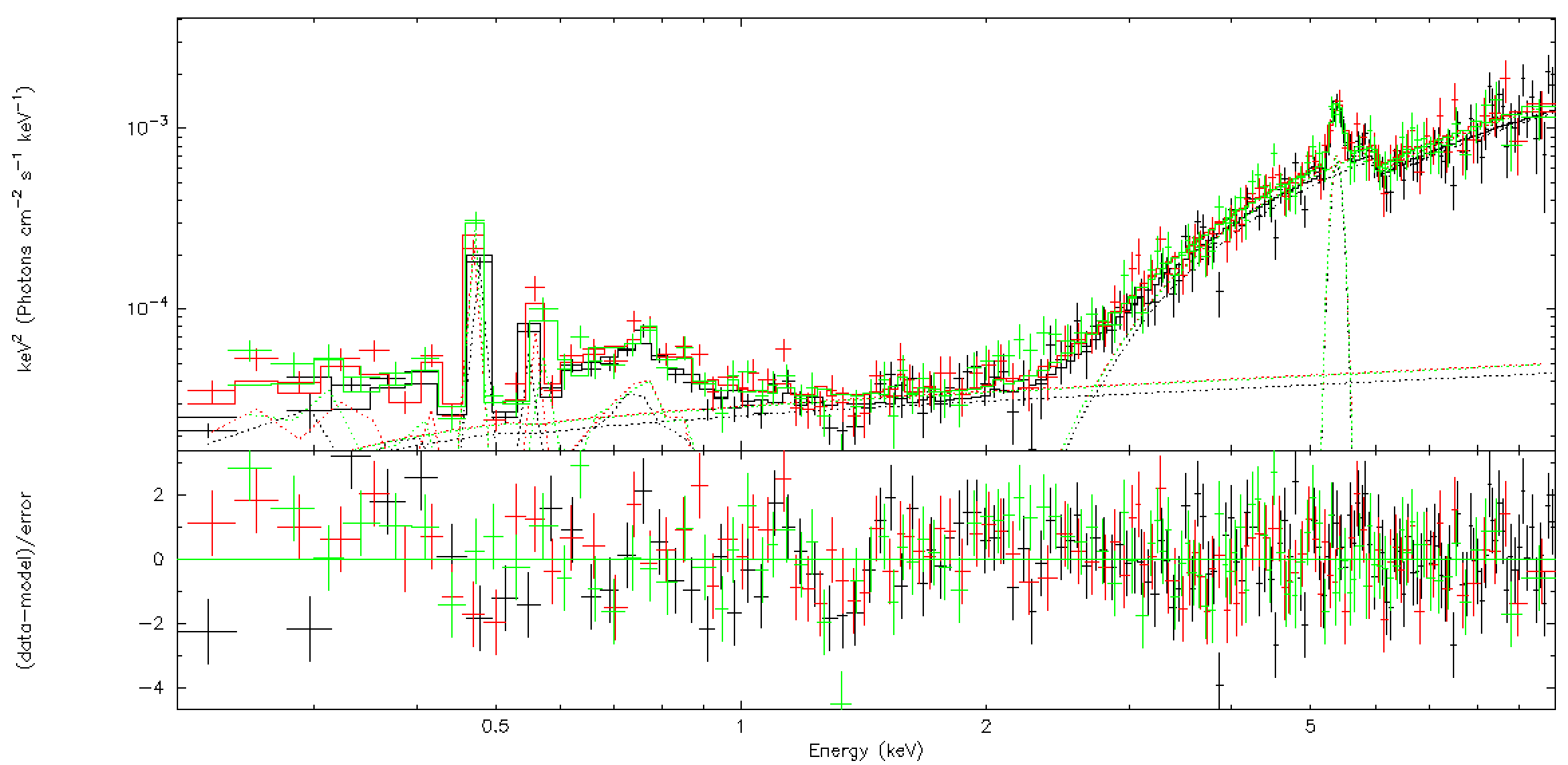}
\caption{{\bf Upper:} The result for the fitting of 3C 234.0. The datapoints from the PN camera are in~black, while the points from the two MOS cameras are in green and red. {\bf Lower:} The residuals of the three fitting procedures as the difference between the model and the~datapoints. \label{fig0}}
\end{figure}

However, in~the case of 3C 234.0, fitting complications arose, in~particular at energies <1 keV, see Figure~\ref{fig0}. The~residuals showed a significant deficit that it was not possible to eliminate introducing an intrinsic absorption. Furthermore, these residuals showed a peculiar trend made by several bumps. To~address these complexities, we utilized the following model: {\tt tbabs*(zpo+apec+apec+ztbabs*(zpo+zga))} similar to those introduced by~\cite{Piconcelli}. This model comprises two components representing collisionally ionized diffuse gas ({\tt apec}) to reproduce the main bumps (see the region around 0.5 keV in Figure~\ref{fig0}), a~model for obscured-type emissions ({\tt tbabs*(zpo+ ztbabs*zpo)}), and~a Gaussian line to account for the iron emission line ({\tt zgauss}).The second part of the model is analogous to the one described earlier and represents the absorbed and scattered emission of the AGN, while the various bumps at <1 keV present a fitting challenge. In~\cite{Piconcelli}, the~authors propose two models to produce a satisfactory result: two components of ionized diffuse gas ({\tt mekal}, similar to {\tt apec}), or~a series of narrow-Gaussian lines. For~our purposes, a~complete physical description of the sources is not needed, so we decided to use the simplest of the two and obtain a reliable measure of the $N_H^{\mathrm{int}}$ parameter. 

For a complete description of the models used, see Tables~\ref{taba2} and~\ref{taba3} and the HEASARC manual\footnote{\url{https://heasarc.gsfc.nasa.gov/xanadu/xspec/manual/Models.html}}.

\unskip

\section{Results}\label{sec3}
In this section, we present our findings in terms of various optical parameters plotted against the X-ray counterparts, notably the total column density as a sum of the galactic one and the intrinsic one $N_H^{\mathrm{tot}}=N_H^{\mathrm{int}}+N_H^{\mathrm{gal}}$. This choice is related to the possibility of comparing our results with other studies and the intent to not pile up all the points with $N_H^{\mathrm{int}}=0$ along the x-axis. Nevertheless, we also generated identical plots excluding the sources with $N_H^{int}$ = 0, see Figures~\ref{figa1} and~\ref{figa2} in Appendix \ref{app2}. This was done to assess whether the relationships with optical estimations of obscuration show improvement. This analysis corresponds to the one conducted in \cite{DallaBarba};
for reference, refer to Figure~6 and Section~5 of \cite{DallaBarba}. 

Figures~\ref{fig1a}--\ref{fig1c} depict analogous plots to those in Figure~6 of \cite{DallaBarba}, where we related the Seyfert type (calculated through the methods presented in~\cite{Whittle,Netzer}) with the Balmer decrement and the [O III] luminosity. Our objective is to identify a correlation between the parameters that demonstrates a differentiation between the jetted and non-jetted sources. Similarly to our previous study, Figures~\ref{fig1a}--\ref{fig1c} represent the relation between $N_H^{\mathrm{tot}}$ and the extinction $A(V)$, the~Seyfert type according to~\cite{Whittle}, and~the [O III] luminosity. All these plots suggests that a distinction between jetted and non-jetted sources is not evident. The~existence of separation would imply a relationship between the nature of the source and the involved~quantities.

Moving to Figure~\ref{fig2}, it showcases the comparison between optical and X-ray luminosities. Although~the relationship between these two quantities is present, it is not particularly strong (see the next sections for the details). These results align with the findings from~\cite{Panessa,Stern}.

\subsection{A(V) versus $N_H^{\mathrm{tot}}$}
\begin{figure}[h]
\hspace{-7mm}
\includegraphics[width=14 cm]{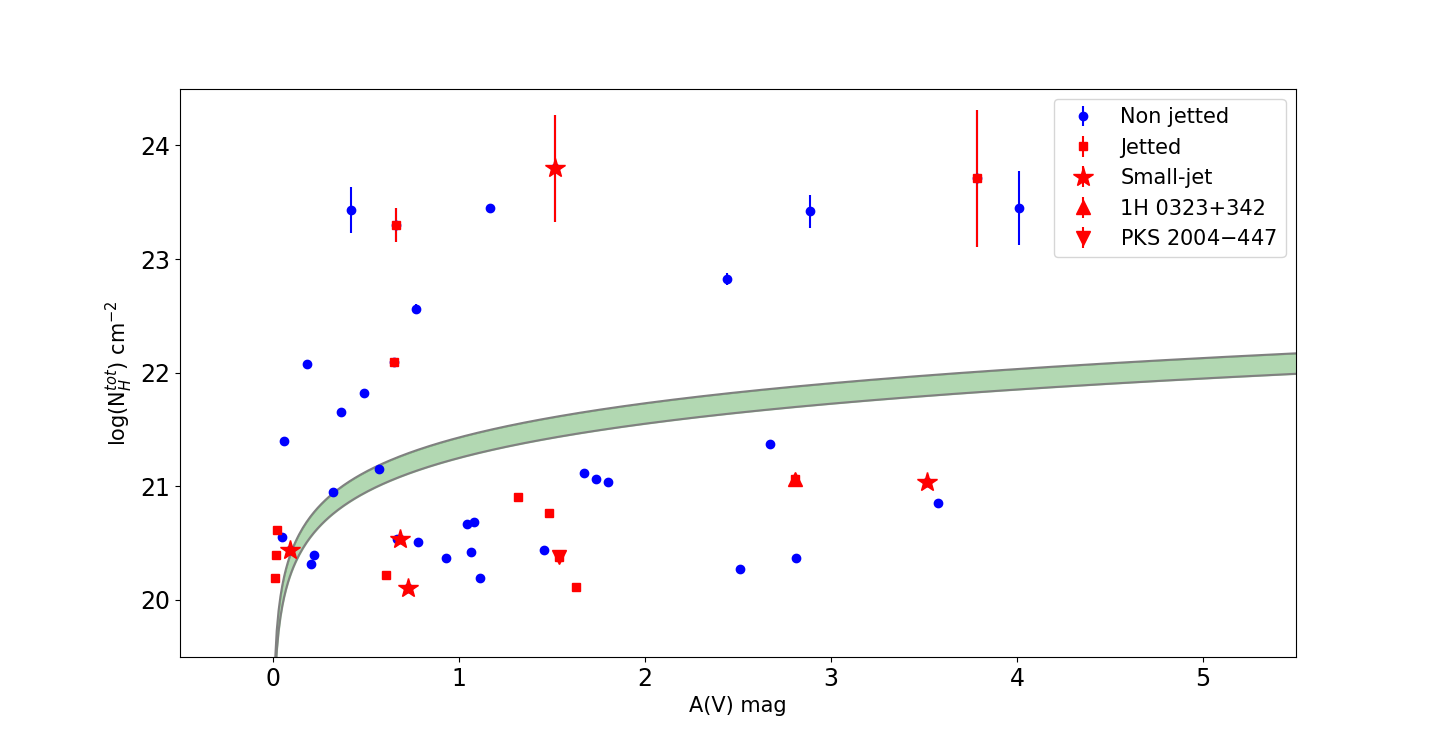}
\caption{The relationship between the hydrogen abundance $N_H^{\mathrm{tot}}$ and the extinction parameter $A(V)$. The~green area marked, sourced from~\cite{Predehl,Nowak}, refers to the Galactic standard~ratio. \label{fig1a}}
\end{figure}

The parameter $A(V)$ quantifies the extinction in magnitudes and directly relies on the Balmer decrement, serving as another means to estimate it. These two parameters are interrelated through the equation provided in~\cite{CCM}:
\begin{equation}
	A(V)=7.215\cdot \mathrm{log}\bigg(\frac{2.86}{\mathrm{H}\alpha/\mathrm{H}\beta}\bigg)
\end{equation}
Here, H$\alpha$ and H$\beta$ represent the flux of the narrow component of the respective lines. The~green area in Figure~\ref{fig1a}, sourced from~\cite{Predehl,Nowak}, reflects the Galactic standard ratio and helps in comparing similar plots, such as the one found in Figure~3 of~\cite{Burtscher}. Notably, in~the latter study, the~authors present a broader extinction range between 0 and 30 mag, whereas in our dataset, the~objects fall within 0 $\leq A(V)<$ 4.5 mag. In this case, there is no need to produce the same plots by discarding objects with $N_H^{int}$ = 0, as~the Galactic contribution is emphasized through the green area.
 
In our study and in~\cite{Burtscher}, the sources seem to scatter around the Galactic ratio within the initial extinction range (e.g., $A(V)<5$ mag). A~similar trend is also noted in another study (Figure 4 of~\cite{Schnorr}), where $A(V)$ ranges from 0 to 10 mag. However, the~limited range of extinction in our dataset limits the possibility of completely comparing our results with previous works. From~the UM and the selection criteria cited in Section~\ref{sec2.1}, we will expect a separation between jetted and non-jetted objects, where the red points concentrate in the low-extinction region (right part of the plot), and~the blue points in the high-extinction one (left side). Though~the sources do not exhibit a clear separation according to this classification, this discrepancy with the UM might partially stem from the restricted range of extinction values available for the~analysis.
\vspace{-12pt}
\unskip

\subsection{[O III]/H$\beta$ versus $N_H^{\mathrm{tot}}$}\label{sec3.2}
The parameter $R$=[O III]/H$\beta$ emerges as a better indicator for determining Seyfert types compared to the H$\alpha$-based method, even at lower types like Sy1.2/Sy1.0, as~highlighted in Figure~6 of \cite{DallaBarba}. In~that plot, the~relationship between the two approaches holds for Seyfert types >Sy1.5, while the scheme proposed by~\cite{Netzer} tends to over-classify Sy1.0/1.2. 

\begin{figure}[h]
\hspace{-7mm}
\includegraphics[width=14 cm]{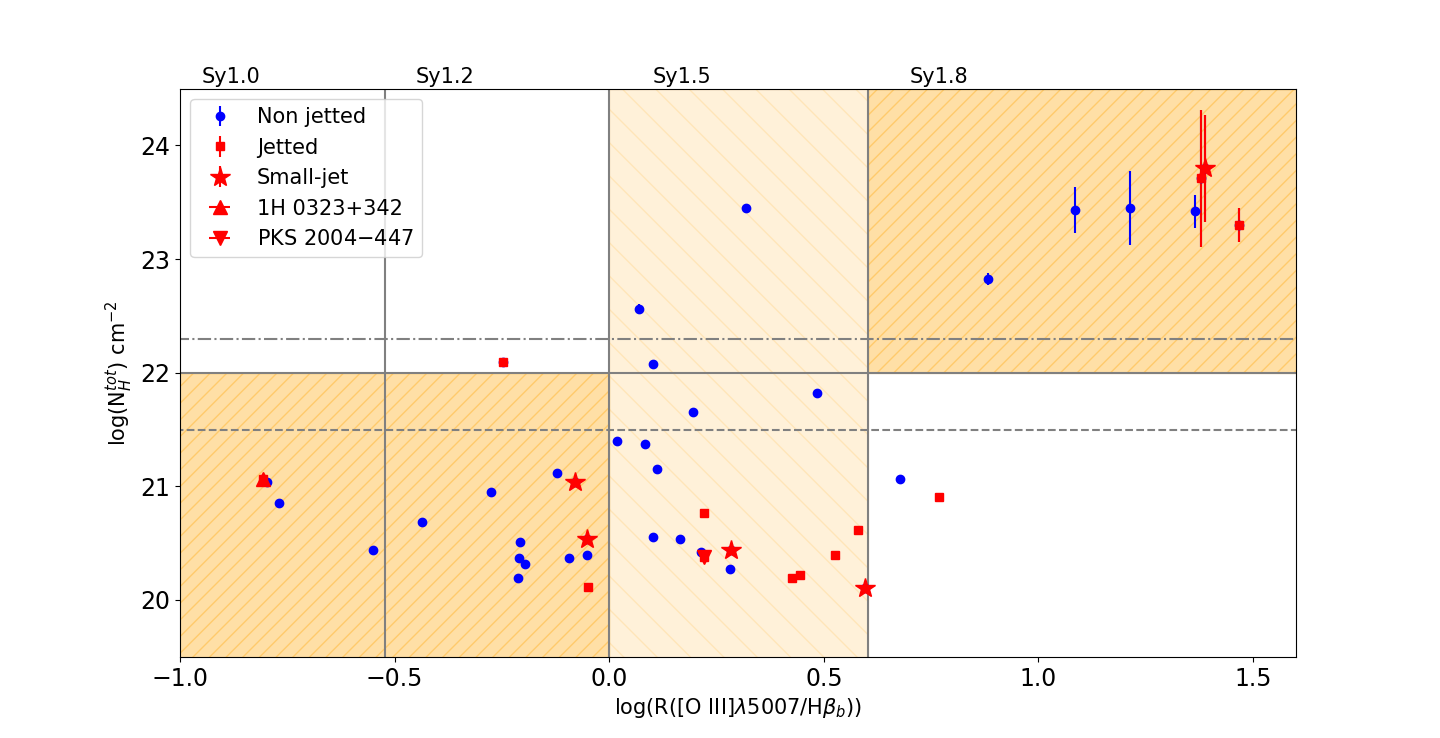}
\caption{The Seyfert type based on the [O III]/H$\beta$ ratio against the hydrogen abundance. A~solid horizontal grey line distinguishes between unobscured and obscured objects, using a threshold of 10$^{22}$ atoms cm$^{-2}$. Additional lines---dashed~\cite{Merloni} and dashed--dotted~\cite{Burtscher}---serve similar purposes. Vertical lines separate Seyfert sub-types according to Equation~(1) and Table~1 in \cite{DallaBarba}. Yellow areas enhance the~separation. \label{fig1b}}
\end{figure}

In the UM, the~separation between obscured and unobscured AGN is typically indicated by an $N_H^{\mathrm{tot}}$ value around $10^{22}$ atoms cm$^{-2}$, as~shown by the horizontal solid line in  Figure~\ref{fig1b}. However, discriminant parameters can vary across different studies, for~instance, using the dashed line from~\cite{Merloni} or the dashed--dotted line from~\cite{Burtscher}. The~vertical lines in the plot indicate Seyfert sub-classes based on the methodology presented in~\cite{Whittle}. 

As anticipated, the~objects concentrate within two yellow regions: one with \mbox{log($N_H^{\mathrm{tot}}$) < 22}, $R$ < 1 and another with log($N_H^{\mathrm{tot}}$) > 22, $R$ > 4. The~faint yellow region corresponds to Sy1.5 (\mbox{1 < $R$ < 4}), and~it is interpreted as a transition zone between obscured and unobscured sources. A~similar analysis was conducted by the authors of~\cite{Burtscher} in their Figure~2, although~their horizontal axis is more refined due to a grouping process into the main sub-types. Nevertheless, our findings align with those reported in~\cite{Burtscher,Risaliti99}, confirming an ascending trend between the Seyfert type and the column density. In~the latter study, the~authors demonstrated that Sy1.8/1.9 typically exhibit log($N_H^{\mathrm{tot}}$) $\sim$ 23--24, while the distribution for Sy2.0 is centered at higher absorption columns (see Figure~5 of~\cite{Risaliti99}).

Figure~\ref{fig1b} reveals that the majority of jetted sources reside in the lower-left part while non-jetted sources occupy both the upper-right and the lower-left part of the figure. This result partially aligns with the UM. However, three jetted objects lie in the high-obscuration region. Their estimation of the $N_H^{\mathrm{tot}}$ parameter is not as reliable as in other cases, possibly due to low-resolution X-ray spectra. These objects include 4C $+29.30$ and 3C 234.0, both radio sources, the~latter involving a complicated fitting process (refer to Section~\ref{sec2.2}), and~2MASX J$04234080+0408017$ with small radio jets~\cite{Smith}. For~radio-emitting objects, significant obscuration is not contradictory to the UM. Indeed, radio emission can be divided into two components: core emission, which is enhanced along the observer's line of sight, akin to $\gamma$-rays; and extended emission, often called lobe emission, which is not beamed. Furthermore, some authors have~\cite{Berton21} stressed the limits in classifying the AGN adopting a radio-loud/radio-quiet scheme. In~this view, there is a continuum between the two classes which can explain the ambiguity of some objects in Figure~\ref{fig1b}. The~non-jetted sources in the lower-left region do not pose a significant challenge to the UM, as~not all poorly-obscured sources necessarily exhibit jets. This ambiguity stems from the selection method adopted, biased toward lower Seyfert types for the jetted group and higher Seyfert types for the non-jetted group, but~this does not result in a distinct~separation.

Even if it is not possible to find a clear division between the two identified classes in this study, our findings broadly concur with previous analyses. Furthermore, eliminating sources with zero intrinsic absorption yields analogous results. The~only noticeable difference is the absence of objects in the Sy1 class. This suggests that the Galactic contribution to the obscuration does not significantly alter the conclusions. See Figure~\ref{figa2} in Appendix \ref{app2} for the comparison plots.

The statistical tests performed result in the following: Spearman ($r_s$ = 0.31, \emph{p}-value = 3.56$\times$10$^{-2}$), Pearson ($r_p$ = 0.55, \emph{p}-value = 5.25$\times$10$^{-5}$), and~Kendall ($r_k$ = 0.20, \emph{p}-value = 4.36$\times$10$^{-2}$). 
\vspace{-12pt}
\unskip

\subsection{L$_{\mathrm{[O III]}}$ versus $N_H^{\mathrm{tot}}$}\label{sec3.3}
\begin{figure}[h]
\hspace{-7mm}
\includegraphics[width=14 cm]{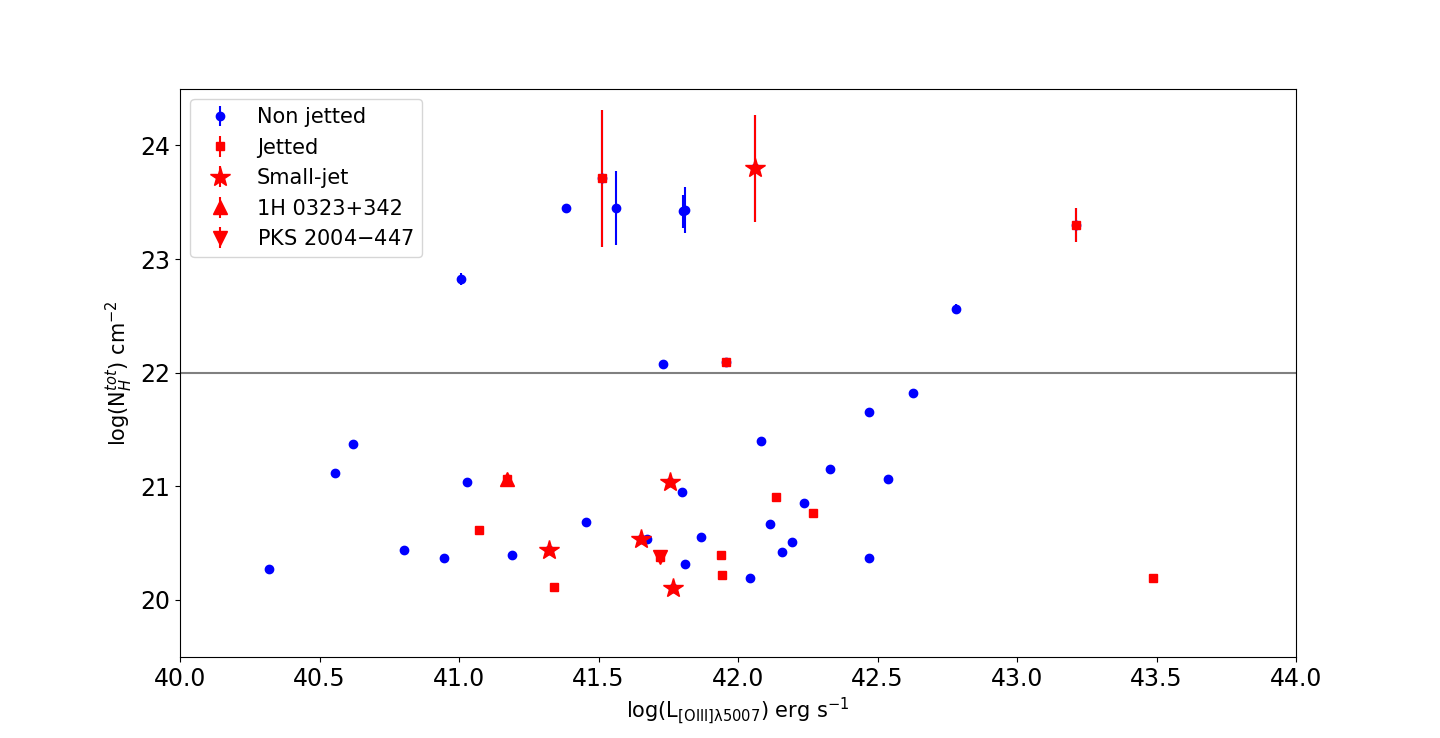}
\caption{The comparison between the [O III]$\lambda$5007 luminosity and the column density. Similar to Figure~\ref{fig1b}, a~horizontal line denotes the separation highlighted in the central panel, likely using the same threshold~criteria. \label{fig1c}}
\end{figure}

The relationship between the [O III] luminosity (L$_{\mathrm{[O III]}}$) and the column density aims to compare the intrinsic properties of AGN, represented by L$_{\mathrm{[O III]}}$ (as detailed in \cite{DallaBarba}), and~obscuration, $N_H^{\mathrm{tot}}$. Comparing Figure~\ref{fig1c} with the similar panel in the previous study, it is evident that the results obtained from optical-based methods align closely with those derived from X-ray-based approaches. In~both cases, the~objects are distributed without a distinct separation between the classes. Similar to the previous section, the~exclusion of objects with $N_H^{int}$ = 0 does not impact the results; in this case as well, the~sources tend to mix and do not exhibit a clear separation between the classes. See Figure~\ref{figa2} in \mbox{Appendix \ref{app2}} for the details.

Consequently, the~orientation-dependent explanation proposed by the UM remains the most viable solution, given the lack of substantial correlation between the intrinsic properties of AGN and obscuration levels across different~classes.

The statistical tests give the following results: Spearman ($r_s$ = 0.08, \emph{p}-value = 5.91$\times$10$^{-1}$), Pearson ($r_p$ = 0.11, \emph{p}-value = 4.78$\times$10$^{-1}$), and~Kendall ($r_k$ = 0.06, \emph{p}-value = 5.57$\times$10$^{-1}$). 
\vspace{-12pt}
\unskip

\subsection{Luminosities}
\begin{figure}[h]
\hspace{-7mm}
\includegraphics[width=14 cm]{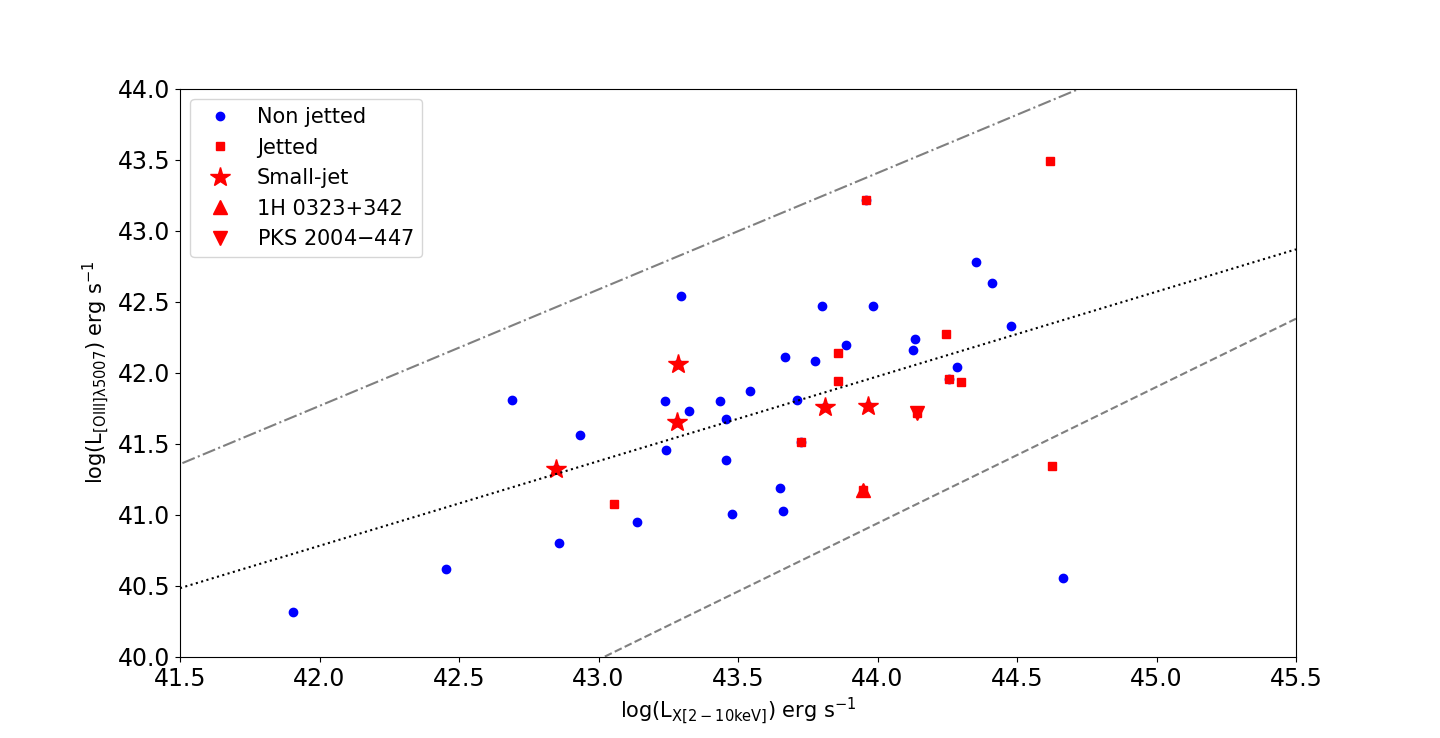}
\caption{The plot compares the X-ray luminosity with the oxygen luminosity. The~dotted black line represents the linear regression line based on our dataset, while the dashed line corresponds to the regression line calculated by~\cite{Stern}. Additionally, the~dashed--dotted line reflects the regression analysis conducted by~\cite{Panessa}. \label{fig2}}
\end{figure}

The plot reported in Figure~\ref{fig2} compares the X-ray luminosity between 2 and 10 keV (L$_{\mathrm{X[2-10keV]}}$) with L$_{\mathrm{[O III]}}$. Within~the same figure, three reference lines are presented: our dataset's linear regression with the dotted black line, the~relation from~\cite{Stern} indicated by the dashed grey line, and~the dashed--dotted line from~\cite{Panessa}. 

Previous studies~\cite{Heckmann} explored the L$_X$-L$_{\mathrm{[O III]}}$ comparison, attempting to distinguish between Sy1 and Sy2. However, even in their analysis, the~objects showed a mix within the two classes. These authors emphasized the selection-dependent nature of this relation, noting its robustness for hard-X-ray-selected objects but its weaker correlation for optically selected AGN. In~our study, where the objects are hard-X-ray-selected, we support the initial findings of~\cite{Heckmann}. 

Focusing on the steepness of the correlations observed, we find different relationships: L$_{\mathrm{[O III]}}$ $\propto$ 7.34 L$_X^{0.82}$ for~\cite{Panessa}, L$_{\mathrm{[O III]}}$ $\propto$ $-$1.30 L$_X^{0.96}$ for~\cite{Stern}, and~L$_{\mathrm{[O III]}}$ $\propto$ 15.74 L$_X^{0.60}$ in our case. The~discrepancies between our findings and those of~\cite{Stern} might arise from differences in the nature of the samples studied. In~our dataset, there is a mixture of Seyfert types, whereas~\cite{Stern} focused solely on type 1 AGN. Additionally, their analysis encompassed 3579 AGN, a~considerably larger sample size compared to our roughly \mbox{50 objects}. However, the~sample size difference between~\cite{Panessa} (60 AGN) and our study is not significant. Neither~\cite{Panessa} nor~\cite{Heckmann} observed a clear separation between Sy1 and Sy2, akin to the jetted/non-jetted classification in our case. This lack of distinction might stem from the limited number of sources considered in both~studies.

The two outliers in the bottom part of the plot correspond to 2MASX J$01492228-5015073$ and GB6 J$0937+5008$. The~optical spectrum of the former exhibits irregularities at different wavelengths, which might influence the process. However, it is important to note that this source is not far from the linear regression identified in this study. Regarding the second case, we have a limited number of bins in the X-ray analysis. This situation can potentially lead to a displacement of the datapoints towards either lower or higher values of L$_{\mathrm{X[2-10keV]}}$, due to the challenge in accurately constraining the parameter within the analysis~process. 

Nevertheless, despite these variations, there is a consensus among these studies about the presence of a relationship between the two luminosities involved, emphasizing the correlation between X-ray and [O III] luminosities across different samples of~AGN.

{The results for the statistical tests performed are as follows: Spearman ($r_s$ = 0.52, \mbox{\emph{p}-value = 1.52$\times$10$^{-4}$}), Pearson ($r_p$ = 0.54, \emph{p}-value = 8.15$\times$10$^{-5}$), and~Kendall ($r_k$ = 0.40, \emph{p}-value = 6.14$\times$10$^{-5}$). }
\vspace{-12pt}
\unskip

\section{Summary and~Conclusions}\label{sec4}
This work serves as the continuation of \cite{DallaBarba}, focusing on the X-ray spectral analysis after analyzing the optical spectra of a combined sample of hard-X-ray- and $\gamma$-ray-selected sources. Initially, our goal was to establish a connection between the intrinsic properties of IS and the inclination-dependent view proposed by the UM. In~\cite{DallaBarba}, we utilized the Balmer decrement and the Seyfert type derived from the H$\beta$/[O III] ratio introduced by~\cite{Whittle} to estimate the obscuration. However, these attempts did not reveal a significant trend or distinct separation between the two classes of~sources. 

To deepen this analysis, we turned to the X-ray spectra to derive a more reliable measure of obscuration represented by $N_H^{\mathrm{tot}}$. The~results, illustrated in Figure~\ref{fig1c}, mirror the optical-based analysis from Figure~6 of \cite{DallaBarba}. This further confirms the absence of a direct relation between the intrinsic properties of IS and obscuration. To further validate this result, we attempted to exclude objects with $N_H^{int}$ = 0 from the sample, considering only those sources exhibiting an intrinsic component of obscuration. Nevertheless, all the conclusions drawn in the first case apply to this second scenario as well, emphasizing that the Galactic contribution is marginal.

In  Figure~\ref{fig1b}, we observe a separation of IS types based on an $N_H^{\mathrm{tot}}$ threshold around 10$^{22}$ atoms cm$^{-2}$ with a transition region corresponding to Sy1.5. This supports the UM's perspective, wherein an increase in Seyfert type correlates with a gradual increase in obscuration, particularly when also considering~ISs.

Moving to Figure~\ref{fig1a}, comparing optical extinction with column density, our analysis is constrained due to the limited horizontal axis. Consequently, our findings lack substantial data for objects with significant extinction. However, they concur with~\cite{Burtscher}, where low-extinction points are not significantly distributed above the Galactic reference ratio in~green. 

In Figure~\ref{fig2}, depicting X-ray and oxygen luminosity, our observations display a slightly different trend compared to~\cite{Panessa,Stern}. This difference in trend might be attributed to variations in the number or nature of sources analyzed across the three studies. Nonetheless, it reinforces the interpretation of oxygen luminosity as a measure of disk luminosity and, by~extension, intrinsic properties, supporting the notion that the X-ray-emitting region is central and closely linked to the accreting~flow. 

In summary, our comprehensive multi-wavelength analysis does not offer clear evidence linking IS type to accretion flow properties. However, we confirm the connection between L$_X$-L$_{\mathrm{[O III]}}$ and the differentiation between Seyfert types and column~density.

ISs might play a pivotal role in understanding other Seyfert families, but~the analysis of large sample sizes can dilute interesting variations between individual objects. This suggests that a more detailed study of individual sources could provide a more robust investigation into the nature of ISs as detailed by~\cite{Jarvela20,Komossa}. \\

\noindent {\textit{Author contribution:} Conceptualization, B.D.B.; Formal analysis, B.D.B.; Supervision, L.F. and M.B.; Validation, L.F., M.B., L.C. and A.V.; Visualization, L.C. and A.V.; Writing---original draft, B.D.B.; Writing---review and editing, L.F., M.B., L.C. and A.V. All authors have read and agreed to the published version of the manuscript. \\

{\small \textit{Acknowledgments:} SDSS acknowledges support and resources from the Center for High-Performance Computing at the University of Utah. The~SDSS web site is \url{www.sdss.org}. SDSS is managed by the Astrophysical Research Consortium for the Participating Institutions of the SDSS Collaboration, including the Carnegie Institution for Science, Chilean National Time Allocation Committee (CNTAC) ratified researchers, the~Gotham Participation Group, Harvard University, Heidelberg University, The Johns Hopkins University, L’Ecole polytechnique f\'{e}d\'{e}rale de Lausanne (EPFL), Leibniz-Institut für Astrophysik Potsdam (AIP), Max-Planck-Institut für Astronomie (MPIA Heidelberg), Max-Planck-Institut für Extraterrestrische Physik (MPE), Nanjing University, National Astronomical Observatories of China (NAOC), New Mexico State University, The Ohio State University, Pennsylvania State University, Smithsonian Astrophysical Observatory, Space Telescope Science Institute (STScI), the~Stellar Astrophysics Participation Group, Universidad Nacional Autonoma de Mexico, University of Arizona, University of Colorado Boulder, University of Illinois at Urbana-Champaign, University of Toronto, University of Utah, University of Virginia, Yale University, and~Yunnan University. This research has made use of the NASA/IPAC Extragalactic Database (NED), which is operated by the Jet Propulsion Laboratory, California Institute of Technology, under~contract with the National Aeronautics and Space Administration. This research has made use of the SIMBAD database, operated at CDS, Strasbourg, France.}}

\begin{landscape}
\appendix
\section[\appendixname~\thesection]{Appendix}\label{app1}
In Table~\ref{taba1}, we reported the list of sources and their main characteristics with the observation IDs. In~Table~\ref{taba2}, we reported the fitting~parameters.

\begin{table}[h]
\small
\caption{The list of sources with the coordinates (RA, DEC) in deg at J2000, the~redshift (z) expressed in terms of $\times10^{-3}$, the~column density ($N_H$) in $\times10^{20}$ cm$^{-2}$, the~observation IDs, the~satellite from which the observations were carried (for {\it XMM-Newton} we use XMM as abbreviation), the~exposure time in ks, and~the starting observation~date.  \label{taba1}}
\newcolumntype{C}{>{\centering\arraybackslash}X}
\begin{tabularx}{\linewidth}{p{2cm}p{4.5cm}p{1.5cm}p{1.5cm}p{1.5cm}p{1cm}p{2cm}p{1.5cm}p{1.5cm}p{2cm}}
\hline
\textbf{IAU Name} & \textbf{Alias} & \textbf{RA} & \textbf{DEC} & \textbf{z} & \textbf{\boldmath$N_H$} & \textbf{Obs. ID(s)} & \textbf{Satellite} & \textbf{Exposure} & \textbf{Date}\\
\hline
\multicolumn{10}{c}{\textbf{BASS}} \\
\hline
J$0149-5015$      &      2MASX   J$01492228-5015073$    &   27.34   &  $-$50.25   &   2.99   &   1.75   &   00038018001   &   {\it Swift}   &   10.58   &   2008-09-11 \\
J$0157+4715$      &      2MASX   J$01571097+4715588$   &   29.30   &   47.27   &   4.79   &   11.0   &   00040895001   &   {\it Swift}   &   6.94   &   2010-10-22 \\
        &        &             &             &             &             &   00040895002   &             &   1.61   &   2010-11-19 \\
J$0206-0017$      &      Mrk   1018   &   31.57   &   $-$0.29   &   4.27   &   2.48   &   00035166001   &   {\it Swift}   &   5.25   &   2005-08-05 \\
        &        &             &             &             &             &   00035776001   &             &   4.86   &   2008-06-11 \\
J$0214-0046$     &      Mrk   590    &   33.64   &   $-$767.00   &   2.63   &   2.77   &   00095662033   &   {\it Swift}   &   9.93   &   2021-01-10 \\
         &       &             &             &             &             &   00037590015   &             &   8.29   &   2016-12-13 \\
J$0234-0847$      &      NGC   985   &   38.66   &   $-$8.79   &   4.30   &   3.48   &   00036530005   &   {\it Swift}   &   9.58   &   2008-06-06 \\
        &        &             &             &             &             &   00096449002   &             &   7.99   &   2021-06-27 \\
J$0238-4038$      &      2MASX   J$02384897-4038377$   &   39.70   &   $-$40.64   &   6.13   &   2.06   &   00049419004   &   {\it Swift}   &   4.76   &   2016-03-27 \\
        &        &             &             &             &             &   00037344002   &             &   4.13   &   2008-03-27 \\
        &        &             &             &             &             &   00037344003   &             &   3.08   &   2008-04-04 \\
        &        &             &             &             &             &   00037344001   &             &   2.71   &   2008-03-13 \\
        &        &             &             &             &             &   00049419001   &             &   1.69   &   2013-03-29 \\
        &        &             &             &             &             &   00093020001   &             &   1.12   &   2017-04-13 \\
        &        &             &             &             &             &   00049419002   &             &   0.88   &   2013-12-27 \\
        &        &             &             &             &             &   00049419003   &             &   0.58   &   2015-09-06 \\
        &        &             &             &             &             &   00049419005   &             &   0.57   &   2016-04-01 \\
\hline
\end{tabularx}
\end{table}
\unskip
\begin{table}[h]
\small
\caption{\em{Cont.}\label{taba1}}
\newcolumntype{C}{>{\centering\arraybackslash}X}
\begin{tabularx}{\linewidth}{p{2cm}p{4.5cm}p{1.5cm}p{1.5cm}p{1.5cm}p{1cm}p{2cm}p{1.5cm}p{1.5cm}p{2cm}}
\hline
\textbf{IAU Name} & \textbf{Alias} & \textbf{RA} & \textbf{DEC} & \textbf{z} & \textbf{\boldmath$N_H$} & \textbf{Obs. ID(s)} & \textbf{Satellite} & \textbf{Exposure} & \textbf{Date}\\
\hline
J$0312+5029$      &      2MASX   J$03120291+5029147$   &   48.01   &   50.49   &   6.15   &   3.45   &   00038026002   &   {\it Swift}   &   6.83   &   2008-12-07 \\
        &        &             &             &             &             &   00038026001   &             &   1.62   &   2008-10-11 \\
J$0330+0538$      &      2MASX   J$03305218+0538253$   &   52.72   &   5.64   &   4.58   &   1.16   &   00039819001   &   {\it Swift}   &   8.86   &   2009-07-04 \\
        &        &             &             &             &             &   00039819002   &             &   2.51   &   2009-07-15 \\
         &       &             &             &             &             &   00047706001   &             &   0.56   &   2013-01-03 \\
         &       &             &             &             &             &   00047706002   &             &   0.40   &   2013-01-09 \\
J$0333+3718$     &      2MASX   J$03331873+3718107$   &   53.33   &   37.30   &   5.47   &   1.48   &   00040693001   &   {\it Swift}   &   6.41   &   2011-04-03 \\
        &        &             &             &             &             &   00031490001   &             &   5.02   &   2009-09-16 \\
        &        &             &             &             &             &   00047806001   &             &   1.47   &   2013-01-06 \\
J$0423+0408$      &      2MASX   J$04234080+0408017$   &   65.92   &   4.13   &   4.62   &   12.6   &   13897   &   {\it Chandra}   &   20.06   &   2012-10-20 \\
J$0503+2300$      &      LEDA   097068   &   75.74   &   23.00   &   5.81   &   23.4   &   00037117001   &   {\it Swift}   &   9.26   &   2007-10-17 \\
        &        &             &             &             &             &   00037117002   &             &   0.81   &   2008-02-17 \\
J$0605-2754$      &      2MASX   J$06054896-2754398$   &   91.45   &   -27.91   &   8.98   &   2.64   &   00090155001   &   {\it Swift}   &   3.89   &   2009-07-06 \\
         &       &             &             &             &             &   00040696003   &             &   3.43   &   2011-05-22 \\
         &       &             &             &             &             &   00040696001   &             &   1.95   &   2011-01-30 \\
         &       &             &             &             &             &   00047747001   &             &   1.31   &   2013-06-24 \\
J$0733+4555$      &      1RXS   J$073308.7+455511$   &   113.29   &   45.92   &   14.15   &   7.19   &   00041758001   &   {\it Swift}   &   6.56   &   2010-12-19 \\
        &        &             &             &             &             &   00041758002   &             &   3.23   &   2010-12-28 \\
        &        &             &             &             &             &   00081466001   &             &   1.99   &   2015-12-26 \\
        &        &             &             &             &             &   00047790002   &             &   1.08   &   2015-05-25 \\
        &        &             &             &             &             &   00041758003   &             &   1.00   &   2010-12-29 \\
        &        &             &             &             &             &   00047790001   &             &   0.39   &   2015-05-24 \\
J$0736+5846$      &      Mrk   9   &   114.24   &   58.77   &   3.99   &   4.83   &   00080535001   &   {\it Swift}   &   6.49   &   2013-10-29 \\
        &        &             &             &             &             &   00041759002   &             &   4.80   &   2010-12-17 \\
        &        &             &             &             &             &   00041759005   &             &   2.69   &   2021-09-22 \\
        &        &             &             &             &             &   00041759004   &             &   2.39   &   2010-12-23 \\
        &        &             &             &             &             &   00041759003   &             &   1.81   &   2010-12-19 \\
        &        &             &             &             &             &   00041759001   &             &   1.74   &   2010-12-07 \\
        &        &             &             &             &             &   07021915001   &             &   0.18   &   2019-10-01 \\
J$0742+4948$     &      Mrk   79    &   115.64   &   49.81   &   2.21   &   5.43   &   00037589002   &   {\it Swift}   &   2.92   &   2009-02-26 \\
        &        &             &             &             &             &   00037589003   &             &   2.83   &   2021-09-22 \\
        &        &             &             &             &             &   00037589001   &             &   2.34   &   2009-02-19 \\
        &        &             &             &             &             &   00096121001   &             &   2.17   &   2021-09-01 \\
\hline
\end{tabularx}
\end{table}
\unskip
\begin{table}[h]
\small
\caption{\em{Cont.}\label{taba1}}
\newcolumntype{C}{>{\centering\arraybackslash}X}
\begin{tabularx}{\linewidth}{p{2cm}p{4.5cm}p{1.5cm}p{1.5cm}p{1.5cm}p{1cm}p{2cm}p{1.5cm}p{1.5cm}p{2cm}}
\hline
\textbf{IAU Name} & \textbf{Alias} & \textbf{RA} & \textbf{DEC} & \textbf{z} & \textbf{\boldmath$N_H$} & \textbf{Obs. ID(s)} & \textbf{Satellite} & \textbf{Exposure} & \textbf{Date}\\
\hline
J$0752+1935$      &      2MASX   J$07521780+1935423$   &   118.07   &   19.60   &   11.70   &   4.06   &   00039551001   &   {\it Swift}   &   5.10   &   2010-05-30 \\
        &        &             &             &             &             &   00038041001   &             &   4.20   &   2008-10-31 \\
        &        &             &             &             &             &   00038041004   &             &   2.91   &   2010-05-30 \\
        &        &             &             &             &             &   00038041003   &             &   1.74   &   2009-09-12 \\
        &        &             &             &             &             &   00038041002   &             &   1.37   &   2009-02-24 \\
J$0803+0841$      &      2MASX   J$08032736+0841523$   &   120.86   &   8.70   &   4.68   &   2.87   &   00038042002   &   {\it Swift}   &   6.60   &   2010-09-15 \\
J$0804+0506$      &      Mrk   1210   &   121.02   &   5.11   &   1.35   &   3.86   &   00037233002   &   {\it Swift}   &   9.61   &   2010-09-10 \\
        &        &             &             &             &             &   00080387001   &             &   6.00   &   2012-10-05 \\
        &        &             &             &             &             &   00035588001   &             &   2.33   &   2006-10-06 \\
        &        &             &             &             &             &   00035588003   &             &   2.32   &   2007-02-25 \\
        &        &             &             &             &             &   00035588002   &             &   1.52   &   2007-01-22 \\
        &        &             &             &             &             &   00037233001   &             &   1.26   &   2008-04-29 \\
J$0829+4154$      &      2MASX   J$08294266+4154366$   &   127.43   &   41.91   &   12.61   &   3.57   &   00040937004   &   {\it Swift}   &   4.41   &   2011-01-17 \\
J$0832+3707$      &      FBQS   J$083225.3+370736$   &   128.11   &   37.13   &   9.20   &   3.27   &   00035634002   &   {\it Swift}   &   4.76   &   2007-02-17 \\
        &        &             &             &             &             &   00035634003   &             &   2.51   &   2007-02-24 \\
        &        &             &             &             &             &   00035634004   &             &   1.82   &   2007-05-01 \\
         &       &             &             &             &             &   00035483003   &             &   1.69   &   2021-02-14 \\
        &        &             &             &             &             &   00035483001   &             &   1.55   &   2006-01-08 \\
        &        &             &             &             &             &   00035483002   &             &   1.12   &   2011-09-09 \\
        &        &             &             &             &             &   00035483004   &             &   1.10   &   2021-03-14 \\
        &        &             &             &             &             &   00035634001   &             &   0.96   &   2007-01-08 \\
J$0840+2949$      &      4C   +29.30   &   130.01   &   29.82   &   6.47   &   4.56   &   11688   &   {\it Chandra}   &   125.09   &   2010-02-19 \\
J$0842+0759$      &      2MASX   J$08420557+0759253$   &   130.52   &   7.99   &   13.37   &   5.76   &   00040938004   &   {\it Swift}   &   7.13   &   2011-01-18 \\
        &        &             &             &             &             &   00040938002   &             &   1.14   &   2010-12-31 \\
        &        &             &             &             &             &   00040938003   &             &   0.98   &   2011-01-09 \\
        &        &             &             &             &             &   00040938001   &             &   0.50   &   2010-12-18 \\
J$0843+3549$      &      2MASX   J$08434495+3549421$   &   130.94   &   35.83   &   5.39   &   2.98   &   18143   &   {\it Chandra}   &   25.07   &   2016-12-26 \\
J$0904+5536$      &      2MASX   J$09043699+5536025$  &   136.15   &   55.60   &   3.72   &   2.32   &   00035260002   &   {\it Swift}   &   7.76   &   2006-01-06 \\
        &        &             &             &             &             &   00035260001   &             &   6.08   &   2005-12-15 \\
        &        &             &             &             &             &   07000264001   &             &   0.06   &   2016-01-13 \\
J$0918+1619$      &      Mrk   704   &   139.61   &   16.31   &   2.95   &   2.74   &   00080414001   &   {\it Swift}   &   6.60   &   2014-12-28 \\
        &        &             &             &             &             &   00035590003   &             &   5.60   &   2006-09-28 \\
        &        &             &             &             &             &   00035590002   &             &   2.26   &   2006-06-14 \\
        &        &             &             &             &             &   00031965001   &             &   1.85   &   2011-04-23 \\
\hline
\end{tabularx}
\end{table}
\unskip
\begin{table}[h]
\small
\caption{\em{Cont.}\label{taba1}}
\newcolumntype{C}{>{\centering\arraybackslash}X}
\begin{tabularx}{\linewidth}{p{2cm}p{4.5cm}p{1.5cm}p{1.5cm}p{1.5cm}p{1cm}p{2cm}p{1.5cm}p{1.5cm}p{2cm}}
\hline
\textbf{IAU Name} & \textbf{Alias} & \textbf{RA} & \textbf{DEC} & \textbf{z} & \textbf{\boldmath$N_H$} & \textbf{Obs. ID(s)} & \textbf{Satellite} & \textbf{Exposure} & \textbf{Date}\\
\hline
        &        &             &             &             &             &   00031965003   &             &   1.85   &   2022-04-23 \\
        &        &             &             &             &             &   00035590005   &             &   1.75   &   2007-01-21 \\
        &        &             &             &             &             &   00015881001   &             &   1.60   &   2023-02-20 \\
        &        &             &             &             &             &   00031965002   &             &   1.14   &   2011-04-23 \\
        &        &             &             &             &             &   00080414002   &             &   1.05   &   2021-03-17 \\
        &        &             &             &             &             &   00035590001   &             &   0.67   &   2006-01-06 \\
J$0923+2255$      &      MCG   $+04-22-042$   &   140.93   &   22.91   &   3.30   &   3.60   &   00035263001   &   {\it Swift}   &   9.20   &   2005-12-10 \\
        &        &             &             &             &             &   00080416001   &             &   3.95   &   2012-12-26 \\
        &        &             &             &             &             &   00089106003   &             &   1.93   &   2021-03-20 \\
        &        &             &             &             &             &   00089106001   &             &   1.92   &   2020-06-05 \\
        &        &             &             &             &             &   00089106004   &             &   1.68   &   2021-05-30 \\
         &       &             &             &             &             &   00089106002   &             &   1.66   &   2020-10-20 \\
        &        &             &             &             &             &   00035263002   &             &   1.02   &   2019-12-25 \\
J$0925+5219$      &      Mrk   110   &   141.30   &   52.29   &   3.52   &   1.27   &   00037561001   &   {\it Swift}   &   10.42   &   2010-01-06 \\
        &        &             &             &             &             &   03111716005   &             &   4.68   &   2022-09-15 \\

        &        &             &             &             &             &   00092396006   &             &   4.12   &   2016-04-29 \\
        &        &             &             &             &             &   00092396007   &             &   4.10   &   2016-04-30 \\
J$0926+1245$      &      Mrk   705  &   141.51   &   12.73   &   2.86   &   3.43   &   00090998001   &   {\it Swift}   &   1.77   &   2011-05-24 \\
        &        &             &             &             &             &   00085341001   &             &   1.01   &   2015-01-02 \\
        &        &             &             &             &             &   00085341004   &             &   0.91   &   2015-04-05 \\
        &        &             &             &             &             &   00085341002   &             &   0.90   &   2015-01-17 \\
J$0935+2617$      &      2MASX   J$09352707+2617093$   &   143.86   &   26.29   &   12.21   &   1.56   &   00040947001   &   {\it Swift}   &   5.82   &   2011-01-02 \\
        &        &             &             &             &             &   00040947003   &             &   3.24   &   2011-01-12 \\
        &        &             &             &             &             &   00040947002   &             &   1.57   &   2011-01-11 \\
J$0942+2341$      &      CGCG   $122-055$   &   145.52   &   23.69   &   2.13   &   2.42   &   00090166001   &   {\it Swift}   &   5.37   &   2009-11-25 \\
J$0945+0738$      &      3C   227   &   146.94   &   7.42   &   8.60   &   2.00   &   00080695001   &   {\it Swift}   &   3.17   &   2014-02-20 \\
        &        &             &             &             &             &   00080695002   &             &   2.40   &   2014-02-21 \\
        &        &             &             &             &             &   00080695003   &             &   2.08   &   2014-02-26 \\
J$0959+1302$      &      NGC   3080   &   149.98   &   13.04   &   3.54   &   2.25   &   00081059001   &   {\it Swift}   &   6.42   &   2017-10-21 \\
        &        &             &             &             &             &   00037365004   &             &   4.32   &   2008-10-10 \\
        &        &             &             &             &             &   00037365003   &             &   3.60   &   2008-10-09 \\
        &        &             &             &             &             &   00037365001   &             &   1.49   &   2008-06-14 \\
        &        &             &             &             &             &   00037365002   &             &   0.97   &   2008-06-30 \\
J$1001+2847$      &      3C   234.0    &   150.46   &   28.79   &   18.48   &   1.62   &   0864430101   &   {\it XMM}   &   95.20   &   2021-05-22 \\
\hline
\end{tabularx}
\end{table}
\unskip
\begin{table}[h]
\small
\caption{\em{Cont.}\label{taba1}}
\newcolumntype{C}{>{\centering\arraybackslash}X}
\begin{tabularx}{\linewidth}{p{2cm}p{4.5cm}p{1.5cm}p{1.5cm}p{1.5cm}p{1cm}p{2cm}p{1.5cm}p{1.5cm}p{2cm}}
\hline
\textbf{IAU Name} & \textbf{Alias} & \textbf{RA} & \textbf{DEC} & \textbf{z} & \textbf{\boldmath$N_H$} & \textbf{Obs. ID(s)} & \textbf{Satellite} & \textbf{Exposure} & \textbf{Date}\\
\hline
J$1023+1951$      &      NGC   3227    &   155.88   &   19.87   &   0.33   &   1.86   &   00092212006   &   {\it Swift}   &   4.16   &   2015-05-09 \\
        &        &             &             &             &             &   00092212011   &             &   3.96   &   2015-10-25 \\
        &        &             &             &             &             &   00092212013   &             &   3.91   &   2015-11-08 \\
        &        &             &             &             &             &   00092212016   &             &   3.87   &   2015-11-29 \\
        &        &             &             &             &             &   00092212014   &             &   3.68   &   2015-11-15 \\
J$1043+1105$      &      SDSS   J$104326.47+110524.2$  &   160.86   &   11.09   &   4.77   &   2.38   &   00040954001   &   {\it Swift}   &   9.95   &   2010-10-29 \\
        &        &             &             &             &             &   00081072001   &             &   5.68   &   2016-06-14 \\
         &       &             &             &             &             &   00088253001   &             &   4.79   &   2019-05-06 \\
J$1315+4424$      &      UGC   08327   &   198.82   &   44.41   &   3.55   &   1.68   &   00037093001   &   {\it Swift}   &   11.49   &   2007-06-03 \\
        &        &             &             &             &             &   00080109001   &             &   6.64   &   2013-04-21 \\
        &        &             &             &             &             &   00037093003   &             &   2.45   &   2007-09-20 \\
        &        &             &             &             &             &   00080109002   &             &   2.01   &   2013-11-18 \\
        &        &             &             &             &             &   00080109003   &             &   1.94   &   2013-11-23 \\
        &        &             &             &             &             &   00088636002   &             &   1.62   &   2017-11-19 \\
        &        &             &             &             &             &   00037093002   &             &   1.23   &   2007-09-19 \\
        &        &             &             &             &             &   00088636001   &             &   0.31   &   2017-11-15 \\
J$1445+2702$      &      CGCG   $164-019$  &   221.40   &   27.03   &   2.96   &   2.37   &   0865140901   &   {\it XMM}   &   27.00   &   2021-01-19 \\

J$1508-0011$      &      Mrk   1393   &   227.22   &   -197.00   &   5.44   &   4.64   &   00030309001   &   {\it Swift}   &   9.76   &   2005-09-10 \\
        &        &             &             &             &             &   00088254003   &             &   7.11   &   2019-03-02 \\
        &        &             &             &             &             &   00088254001   &             &   6.04   &   2018-03-27 \\
        &        &             &             &             &             &   00088254002   &             &   5.76   &   2018-05-13 \\
\hline
\multicolumn{10}{c}{\textbf{4FGL}} \\
\hline
J$0038-0207$      &      3C   17   &   9.59   &   -2.13   &   22.04   &   2.48   &   9292   &   {\it Chandra}   &   8.04   &   2008-02-02 \\
J$0324+3410$      &      1H   0323+342    &   51.17   &   34.18   &   6.29   &   11.7   &   00035630002   &   {\it Swift}   &   8.96   &   2006-07-09 \\
        &        &             &             &             &             &   00035372001   &             &   8.36   &   2006-07-06 \\
        &        &             &             &             &             &   00035630003   &             &   6.87   &   2006-07-14 \\
J$0937+5008$      &      GB6   J$0937+5008$  &   144.30   &   50.15   &   27.55   &   1.31   &   00040543003   &   {\it Swift}   &   2.99   &   2012-09-06 \\
        &        &             &             &             &             &   00040543002   &             &   1.08   &   2011-03-30 \\
        &        &             &             &             &             &   00040543001   &             &   0.85   &   2011-03-30 \\
J$0958+3224$      &      3C   232   &   149.59   &   32.40   &   53.06   &   1.57   &   00010070001   &   {\it Swift}   &   1.96   &   2017-04-29 \\
J$1443+5201$      &      3C   303   &   220.76   &   52.03   &   14.12   &   1.66   &   1623   &   {\it Chandra}   &   15.10   &   2001-03-23 \\
J$1516+0015$      &      PKS   1514+00  &   229.17   &   0.25   &   5.26   &   4.17   &   00088374001   &   {\it Swift}   &   6.53   &   2020-03-15 \\
        &        &             &             &             &             &   00036340002   &             &   6.28   &   2007-06-22 \\
        &        &             &             &             &             &   00036340003   &             &   4.59   &   2008-01-06 \\
\hline
\end{tabularx}
\end{table}
\unskip
\begin{table}[h]
\small
\caption{\em{Cont.}\label{taba1}}
\newcolumntype{C}{>{\centering\arraybackslash}X}
\begin{tabularx}{\linewidth}{p{2cm}p{4.5cm}p{1.5cm}p{1.5cm}p{1.5cm}p{1cm}p{2cm}p{1.5cm}p{1.5cm}p{2cm}}
\hline
\textbf{IAU Name} & \textbf{Alias} & \textbf{RA} & \textbf{DEC} & \textbf{z} & \textbf{\boldmath$N_H$} & \textbf{Obs. ID(s)} & \textbf{Satellite} & \textbf{Exposure} & \textbf{Date}\\
\hline
        &        &             &             &             &             &   00036340006   &             &   2.41   &   2018-12-23 \\
        &        &             &             &             &             &   00036340004   &             &   2.38   &   2010-09-25 \\
        &        &             &             &             &             &   00036340007   &             &   1.25   &   2018-12-26 \\
J$2007-4434$      &      PKS   $2004-447$   &   301.98   &   $-$44.58   &   24.00   &   2.97   &   00032492015   &   {\it Swift}   &   12.26   &   2013-11-20 \\
        &        &             &             &             &             &   00032492006   &             &   11.62   &   2013-07-14 \\
        &        &             &             &             &             &   00032492005   &             &   11.45   &   2013-07-07 \\
        &        &             &             &             &             &   00032492017   &             &   9.70   &   2014-03-16 \\
        &        &             &             &             &             &   00032492007   &             &   8.42   &   2013-09-27 \\
J$2118+0013$      &      PMN   J$2118+0013$   &   319.57   &   0.22   &   46.28   &   5.87   &   00091075002   &   {\it Swift}   &   3.58   &   2011-08-11 \\
        &        &             &             &             &             &   00040665003   &             &   2.69   &   2012-08-26 \\
        &        &             &             &             &             &   00091075006   &             &   2.14   &   2011-08-22 \\
        &        &             &             &             &             &   00091075004   &             &   1.83   &   2011-08-16 \\
        &        &             &             &             &             &   00040665004   &             &   1.56   &   2012-09-02 \\
        &        &             &             &             &             &   00091075001   &             &   0.76   &   2011-05-29 \\
        &        &             &             &             &             &   00040665001   &             &   0.57   &   2011-08-27 \\
        &        &             &             &             &             &   00091075003   &             &   0.48   &   2011-08-15 \\
        &        &             &             &             &             &   00091075005   &             &   0.47   &   2011-08-17 \\
        &        &             &             &             &             &   00040665002   &             &   0.31   &   2011-08-30 \\

J$2118-0732$      &      TXS   $2116-077$    &   319.72   &   $-$7.54   &   26.01   &   7.99   &   00011263004   &   {\it Swift}   &   3.79   &   2019-08-05 \\
        &        &             &             &             &             &   00011263005   &             &   3.79   &   2019-09-05 \\
        &        &             &             &             &             &   00011263003   &             &   3.37   &   2019-07-05 \\
        &        &             &             &             &             &   00011263006   &             &   2.71   &   2019-10-11 \\
        &        &             &             &             &             &   00011263001   &             &   2.55   &   2019-05-05 \\
        &        &             &             &             &             &   00011263002   &             &   2.18   &   2019-06-05 \\
        &        &             &             &             &             &   00011263008   &             &   1.27   &   2019-10-21 \\
        &        &             &             &             &             &   00011263009   &             &   1.21   &   2019-10-22 \\
\hline
\end{tabularx}
\end{table}
\unskip

\begin{table}[h]
\small
\caption{The fitting parameters are as follows: $\Gamma_1$ and $\Gamma_2$ for the power-law indexes, $E_{\mathrm{break}}$ for the break energy for the broken power-law cases in keV, $E_{\mathrm{line}}$ for the Fe K$\alpha$ line energy (when present) in keV, $kT$ for the two {\tt apec} functions used in the fitting of 3C 234.0 in keV, $\chi^2/$dof for the goodness-of-fit in terms of the $\chi^2$ and the degrees of freedom (dof), the~adopted model according to the abbreviations listed in Table~\ref{taba3}, $F^{\mathrm{obs}}_{\mathrm{2-10 keV}}$ for the observed flux in units of 10$^{-12}$ erg cm$^{-2}$ s$^{-1}$, $L^{\mathrm{int}}_{\mathrm{2-10keV}}$ for the intrinsic luminosity removing the absorption components coming from galactic and internal factors in units of 10$^{43}$ erg s$^{-1}$, and~$N_H^{\mathrm{int}}$ for the additional obscuration when present in units of 10$^{20}$ atoms cm$^{-2}$. With~($^a$), we highlight the fixed parameters in the fit and with ($^b$) the results for the fitting of 3C 232 based on the rare-events statistics ({\it c-stat}).\label{taba2}}
\begin{tabularx}{\linewidth}{p{2cm}p{1.5cm}p{1.5cm}p{1.5cm}p{1.5cm}p{1.5cm}p{1.5cm}p{1.25cm}p{1.75cm}p{1.25cm}p{1.25cm}p{2cm}}
\hline
\textbf{IAU Name} & \textbf{\boldmath$\Gamma_1$} & \textbf{\boldmath$\Gamma_2$} & \textbf{\boldmath$E_{\mathrm{break}}$} & \textbf{\boldmath$E_{\mathrm{line}}$} & \textbf{\boldmath$kT_1$} & \textbf{\boldmath$kT_2$} & \textbf{\boldmath$\chi^2/$dof} & \textbf{Model} & \textbf{\boldmath$F^{\mathrm{obs}}_{2-10 \mathrm{keV}}$} & \textbf{\boldmath$L^{\mathrm{int}}_{2-10 \mathrm{keV}}$} & \textbf{\boldmath$N_H^{\mathrm{int}}$}\\
\hline
\multicolumn{12}{c}{\textbf{BASS}} \\	
\hline
J$0149-5015$  &  2.00$_{-0.18}^{0.19}$  &  -  &  -  &  -  &  -  &  -  &  33/38  &  ABS+PL  &  394.05  &  0.99  &  0.11$^{+ 0.04 }_{ -0.04}$   \\
J$0157+4715$  &  2.87$_{-1.25}^{2.86}$  &  1.98$_{-0.06}^{0.06} $ &  1.24$_{-0.07}^{0.16}$  &  -  &  -  &  -  &  113/104  &  PL  &  14.57  &  9.24  &  0.0   \\
J$0206-0017$  &  1.90$^{+0.04}_{-0.04}$  &  -  &  -  &  -  &  -  &  -  &  198/184  &  PL  &  21.07  &  8.88  &  0.0  \\
J$0214-0046$  &  1.61$^{+0.05}_{-0.05}$  &  -  &  -  &  -  &  -  &  -  &  104/111  &  PL  &  7.48  &  1.13  &  0.0\\
J$0234-0847$  &  3.53$^{+0.13}_{-0.12}$  &  1.60$^{+0.04}_{-0.04}$  &  0.94$^{+0.03}_{-0.03}$  &  -  &  -  &  -  &  293/262  &  ABS+BPL  &  25.99  &  0.15  &  7.43$^{+ 0.03}_{ -0.03}$   \\
J$0238-4038$  &  1.79$^{+0.04}_{-0.04}$  &  -  &  -  &  -  &  -  &  -  &  181/180  &  PL  &  10.94  &  9.35  &  0.0  \\
J$0312+5029$  &  1.72$^{+0.17}_{-0.16}$  &  -  &  -  &  -  &  -  &  -  &  52/48  &  ABS+PL  &  9.11  &  0.11  &  0.11$^{+ 0.08}_{ -0.08}$   \\
J$0330+0538$  &  1.71$^{+0.08}_{-0.08}$  &  -  &  -  &  -  &  -  &  -  &  62/59  &  PL  &  6.31  &  3.38  &  0.0  \\
J$0333+3718$  &  1.57$^{+0.10}_{-0.10}$  &  -  &  -  &  -  &  -  &  -  &  87/90  &  ABS+PL  &  11.48  &  9.26  &  0.10$^{+0.05}_{ -0.04}$   \\
J$0423+0408$  &  2.59$^{+0.23}_{-0.22}$  &  1.90 $^a$  &  -  &  -  &  -  &  -  &  61/33  &  OBS  &  1.51  &  8.28  &  62.84$^{+11.22}_{-9.72}$   \\
J$0503+2300$  &  1.23$^{+0.04}_{-0.04}$  &  -  &  -  &  -  &  -  &  -  &  149/149  &  PL  &  19.71  &  0.18  & 0.0   \\
J$0605-2754$  &  1.47$^{+0.05}_{-0.05}$  &  -  &  -  &  -  &  -  &  -  &  81/88  &  PL  &  10.85  &  0.20  &  0.0   \\
J$0733+4555$  &  2.07$^{+0.07}_{-0.07}$  &  -  &  -  &  -  &  -  &  -  &  91/78  &  PL  &  5.19  &  0.32  &  0.0   \\
J$0736+5846$  &  1.99$^{+0.04}_{-0.04}$  &  -  &  -  &  -  &  -  &  -  &  231/170  &  PL  &  10.06  &  3.77  &  0.0  \\
J$0742+4948$  &  2.14$^{+0.12}_{-0.11}$  &  1.31$^{+0.33}_{-0.35}$  &  -  &  -  &  -  &  -  &  240/212 & OBS  &  30.96  &  3.92  &  1.14$^{+ 0.80}_{-0.53}$   \\
J$0752+1935$  &  1.90 $^a$  &  1.51$^{+1.11}_{-0.51}$  &  -  &  -  &  -  &  -  &  32/34  &  OBS  &  6.02  &  0.33  &  3.60$^{+ 1.07}_{-0.89}$   \\
J$0803+0841$  &  1.51$^{+0.14}_{-0.13}$  &  -  &  -  &  -  &  -  &  -  &  62/54  &  ABS+PL  &  7.85  &  4.31  &  0.17$^{+ 0.07}_{ -0.06}$   \\
J$0804+0506$  &  2.29$^{+0.18}_{-0.18}$  &  1.86$^{+0.74}_{-0.66}$  &  -  &  6.25$^{+0.19}_{-0.11}$  &  -  &  -  &  34/41  &  OBS+FE  &  7.54  &  1.50  &  28.26$^{+ 7.65}_{ -6.51}$   \\
J$0829+4154$  &  1.55$^{+0.11}_{-0.10}$  &  -  &  -  &  -  &  -  &  -  &  57/74  &  ABS+PL  &  11.09  &  0.46  &  0.11$^{+ 0.04}_{ -0.04}$   \\
J$0832+3707$  &  1.76$^{+0.07}_{-0.07}$  &  -  &  -  &  -  &  -  &  -  &  85/76  &  PL  &  6.71  &  0.14  &  0.0  \\
J$0840+2949$  &  2.98$^{+0.43}_{-0.42}$  &  1.72$^{+0.93}_{-0.80}$  &  -  &  -  &  -  &  -  &  46/29  &  OBS  &  1.26  &  9.12  &  51.44$^{+ 14.30}_{ -11.34}$   \\
J$0842+0759$  &  1.36$^{+0.21}_{-0.19}$  &  -  &  -  &  -  &  -  &  -  &  24/36  &  ABS+PL  &  6.97  &  0.36  &  0.60$^{+ 0.21}_{ -0.17}$   \\
J$0843+3549$  &  2.42$^{+0.70}_{-0.76}$  &  1.90 $^a$  &  -  &  -  &  -  &  -  &  44/36  &  OBS  &  1.52  &  5.38  &  26.30$^{+ 3.44}_{ -2.90}$   \\
J$0904+5536$  &  1.49$^{+0.06}_{-0.06}$  &  -  &  -  &  -  &  -  &  -  &  97/75  &  BPL  &  7.40  &  2.20  &  0.0  \\
J$0918+1619$  &  3.29$^{+0.38}_{-0.37}$  &  1.34$^{+0.06}_{-0.06}$  &  0.95$^{+0.05}_{-0.05}$  &  -  &  -  &  -  &  251/215  &  ABS+BPL  &  14.54  &  3.14  &  6.10$^{+ 0.03}_{ -0.03}$   \\
J$0923+2255$  &  2.01$^{+0.06}_{-0.05}$  &  1.76$^{+0.04}_{-0.04}$  &  1.29$^{+0.23}_{-0.20}$  &  -  &  -  &  -  &  363/297  &  BPL  &  26.24  &  6.59  &  0.0  \\

\hline
\end{tabularx}
\end{table}
\unskip
\begin{table}[h]
\small
\caption{{\em Cont.}}
\begin{tabularx}{\linewidth}{p{2cm}p{1.5cm}p{1.5cm}p{1.5cm}p{1.5cm}p{1.5cm}p{1.5cm}p{1.25cm}p{1.75cm}p{1.25cm}p{1.25cm}p{2cm}}
\hline
\textbf{IAU Name} & \textbf{\boldmath$\Gamma_1$} & \textbf{\boldmath$\Gamma_2$} & \textbf{\boldmath$E_{\mathrm{break}}$} & \textbf{\boldmath$E_{\mathrm{line}}$} & \textbf{\boldmath$kT_1$} & \textbf{\boldmath$kT_2$} & \textbf{\boldmath$\chi^2/$dof} & \textbf{Model} & \textbf{\boldmath$F^{\mathrm{obs}}_{2-10 \mathrm{keV}}$} & \textbf{\boldmath$L^{\mathrm{int}}_{2-10 \mathrm{keV}}$} & \textbf{\boldmath$N_H^{\mathrm{int}}$}\\
\hline
J$0925+5219$  &  2.09$^{+0.05}_{-0.04}$  &  1.60$^{+0.04}_{-0.04}$  &  1.29$^{+0.13}_{-0.13}$  &  -  &   -  &  -  & 397/373  &  BPL  &  60.55  &  0.16  &  0.0 \\

J$0926+1245$  &  2.05$^{+0.06}_{-0.06}$  &  -  &  -  &  -  &  -  &  -  &  103/82  &  PL  &  22.47  &  4.37  &  0.0 \\
J$0935+2617$  &  1.79 $^a$  &  -  &  -  &  -  &  -  &  -  &  84/98  &  PL  &  9.65  &  0.35  &  0.0 \\
J$0942+2341$  &  1.47$^{+0.13}_{-0.13}$  &  -  &  -  &  -  &  -  &  -  &  30/19  &  ABS+BPL  &  4.23  &  1.17  &  0.21$^{+ 0.13}_{ -0.13}$   \\
J$0945+0738$  &  1.08$^{+0.25}_{-0.24}$  &  -  &  -  &  -  &  -  &  -  &  64/39  &  OBS  &  11.81  &  0.25  &  1.22$^{+ 0.31}_{ -0.26}$   \\
J$0959+1302$  &  1.88$^{+0.07}_{-0.07}$  &  -  &  -  &  -  &  -  &  -  &  74/64  &  PL  &  5.08  &  1.41  &  0.0  \\
J$1001+2847$  &  1.78$^{+0.22}_{-0.27}$  &  1.08$^{+0.18}_{-0.17}$  &  -  &  6.38$^{+0.02}_{-0.02}$  & 0.16$_{-0.01}^{+0.01}$   &  0.81$_{-0.05}^{+0.05}$  &  483/369  &  2APEC+FE  &  1.37  &  0.22  &  20.02$^{+ 3.53}_{ -3.00}$   \\
J$1023+1951$  &  1.01$^{+0.05}_{-0.05}$  &  1.56$^{+0.07}_{-0.06}$  &  1.92$^{+0.24}_{-0.18}$  &  -  &  -  &  -  &  398/364  &  BPL  &  38.11  &  11.25  &  0.0 \\
J$1043+1105$  &  1.63$^{+0.07}_{-0.07}$  &  -  &  -  &  -  &  -  &  -  &  118/137  &  ABS+PL  &  9.00  &  4.67  &  1.05$^{+ 0.02}_{ -0.01}$   \\
J$1315+4424$  &  2.32$^{+0.95}_{-0.91}$  &  1.66$^{+0.29}_{-0.27}$  &  -  &  -  &  -  &  -  &  54/66  &  OBS  &  7.77  &  5.11  &  6.67$^{+ 1.22}_{ -1.02}$   \\
J$1445+2702$  &  2.68$^{+0.10}_{-0.10}$  &  1.74$^{+0.34}_{-0.32}$  &  -  &  -  &  -  &  -  &  173/125  &  OBS  &  1.20  &  1.03  &  26.96$^{+ 4.65}_{ -4.22}$   \\
J$1508-0011$  &  2.33$^{+0.16}_{-0.15}$  &  1.44$^{+0.05}_{-0.05}$  &  0.96$^{+0.11}_{-0.10}$  &  -  &  -  &  -  &  167/168  &  BPL  &  10.81  &  7.48  &  0.0   \\
\hline
\multicolumn{12}{c}{\textbf{4FGL}} \\
\hline
J$0038-0207$  &  1.86$^{+0.42}_{-0.38}$  &  1.27$^{+0.09}_{-0.16}$  &  0.98$^{+0.80}_{-0.26}$  &  -  &  -  &  -  &  77/66  &  BPL  &  2.36  &  0.28  &  0.0  \\
J$0324+3410$  &  3.10$^{+0.50}_{-0.53}$  &  1.83$^{+0.08}_{-0.11}$  &  1.29$^{+0.22}_{-0.07}$  &  -  &  -  &  -  &  261/231  &  PL  &  15.76  &  0.28  &  0.0 \\
J$0937+5008$  &  1.44$^{+0.16}_{-0.16}$  &  -  &  -  &  -  &  -  &  -  &  8/11  &  PL  &  3.02  &  0.61  &  0.0  \\
J$0958+3224$  &  2.03$^{+0.99}_{-0.80}$  &  -  &  -  &  -  &  -  &  -  &  13/22 $^b$ &  PL  &  0.78  &  0.71  &  0.0   \\
J$1443+5201$  &  1.24$^{+0.05}_{-0.05}$  &  -  &  -  &  -  &  -  &  -  &  148/121  &  PL  &  2.11  &  9.43  &  0.0   \\
J$1516+0015$  &  1.57$^{+0.07}_{-0.07}$  &  -  &  -  &  -  &  -  &  -  &  43/65  &  PL  &  2.78  &  1.77  &  0.0  \\
J$2007-4434$  &  1.48$^{+0.07}_{-0.07}$  &  -  &  -  &  -  &  -  &  -  &  46/54  &  PL  &  1.37  &  0.20  &  0.0  \\
J$2118+0013$  &  1.94$^{+0.28}_{-0.27}$  &  -  &  -  &  -  &  -  &  -  &  5/3  &  PL  &  0.43  &  24.74  &  0.0 \\
J$2118-0732$  &  1.51$^{+0.21}_{-0.20}$  &  -  &  -  &  -  &  -  &  -  &  11/8  &  PL  &  0.57  &  0.11  &  0.0  \\
\hline
\end{tabularx}
\end{table}
\unskip

\begin{table}[h]
\caption{List of 
 adopted models and the abbreviations in Table~\ref{taba2}. \label{taba3}}
\begin{tabularx}{\linewidth}{p{8cm}p{8cm}p{8cm}}
\hline
\textbf{Abbreviation} & \textbf{Extended name} & \textbf{Model}\\
\hline
PL & Power-law & {\tt tbabs*zpo} \\
BPL & Broken power-law & {\tt tbabs*bknpo} \\
ABS & Intrinsic absorption & {\tt tbabs*ztbabs*zpo}  \\
 &  & {\tt tbabs*ztbabs*bknpo} \\
OBS & Obscured-type emission & {\tt tbabs*(zpo+ztbabs*zpo)} \\
FE & Iron line & (One of the previous){\tt +zgauss} \\
APEC & Collisionally ionized diffuse gas & (One of the previous){\tt +apec} \\
\hline
\end{tabularx}
\end{table}

\end{landscape}

\section[\appendixname~\thesection]{Appendix}\label{app2}
\begin{figure}[h]
\includegraphics[width=14 cm]{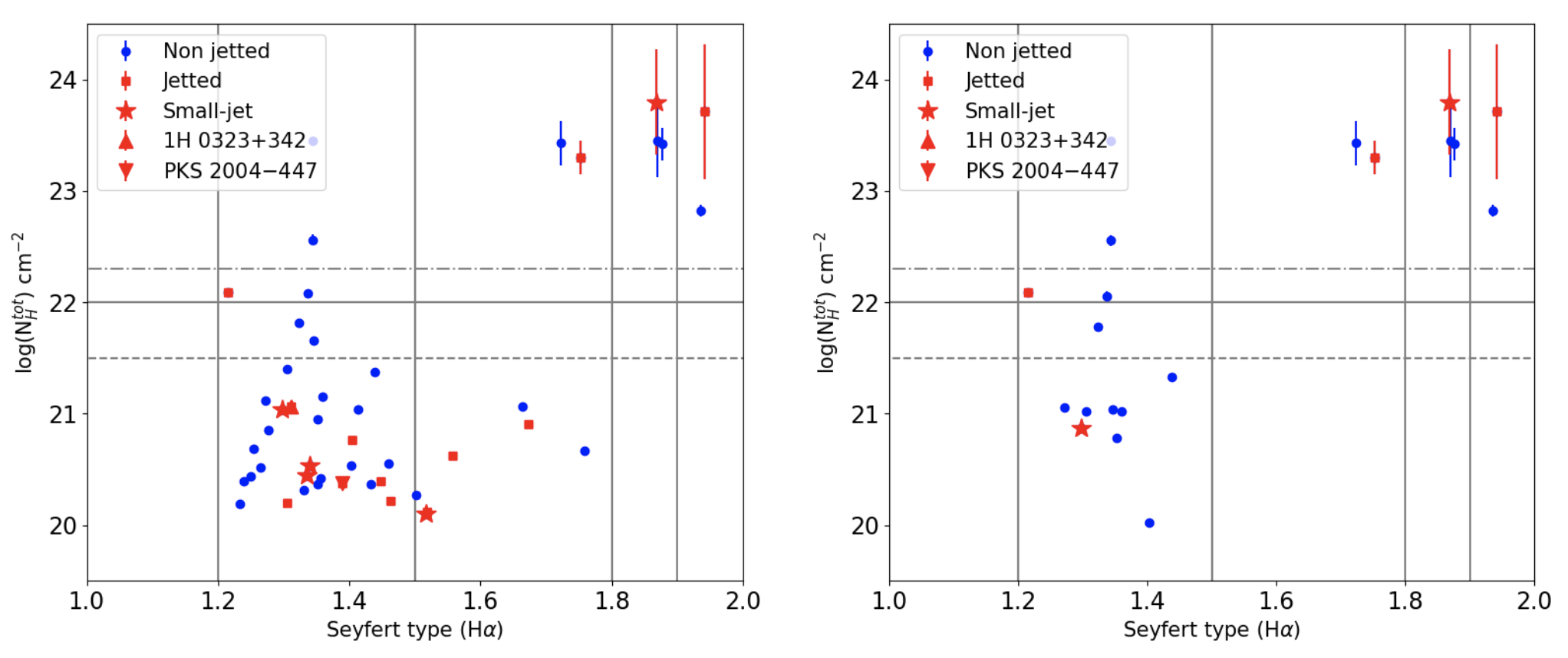}
\caption{Comparison between the Seyfert types using H$\alpha$ and $N_H$ for the complete (\textbf{left}) and the reduced sample (without $N_H^{int}=0$, \textbf{right}). \label{figa1}}   
\end{figure}

In Figure~\ref{figa1}, we provide an alternative representation of Figure~\ref{fig1b}, utilizing the Seyfert type classification determined solely by the properties of the H$\alpha$ line~\cite{Netzer}. However, as~explained in \cite{DallaBarba}, the~method based on H$\beta$/[O III] ratio proposed by~\cite{Whittle} is more effective in classifying objects at lower Seyfert types. This is evident in Figure~\ref{figa1}, where none of the objects fall within the Sy1 region. Notably, even 1H 0323+342 and PKS 2004$-$447, known as two narrow-line Seyfert 1, are not classified within the Sy1 region using the H$\alpha$-based method. The same plot has been generated by excluding objects with $N_H^{int}$ = 0. From~this alternative approach, we can observe that the conclusions drawn in the first case also hold here. 

\begin{figure}[h]
\includegraphics[width=14 cm]{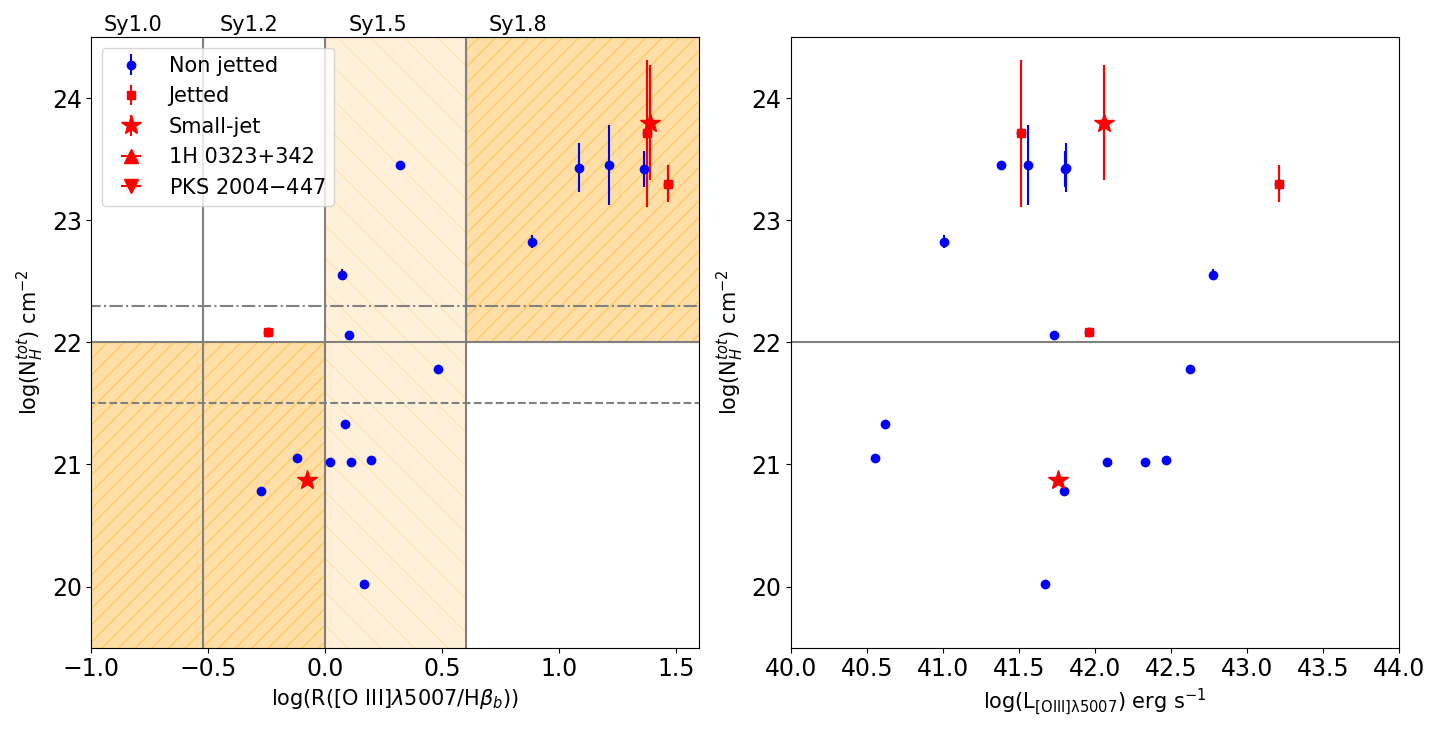}
\caption{The same plots as in Figures~\ref{fig1a} (Seyfert type through [O III]/H$\beta$ versus $N_H$, \textbf{left}) and~\ref{fig1b} (oxygen luminosity versus $N_H$, \textbf{right}) excluding the objects with $N_H^{int}$ = 0. \label{figa2}}   
\end{figure}

\noindent In Figure~\ref{figa2}, we present alternative plots, considering only those objects with $N_H^{int}$ = 0. This serves as a secondary validation of our results, focusing exclusively on the intrinsic characteristics of the selected sources. In~Sections~\ref{sec3.2} and \ref{sec3.3}, we plotted the total $N_H$ parameter, which includes the Galactic component. This may scatter the points as it is not solely dependent on intrinsic obscuration, as~in the optical case. However, as~depicted in the following figure, the~results confirm the earlier conclusions and emphasize the absence of a clear separation between jetted and non-jetted sources. Statistical tests further support this secondary conclusion: Spearman ($r_s$ = 0.73, \emph{p}-value = 2.39$\times$10$^{-4}$), Pearson ($r_p$ = 0.81, \emph{p}-value = 1.23$\times$10$^{-5}$), and~Kendall ($r_k$ = 0.54, \emph{p}-value = 8.27$\times$10$^{-4}$) for the first panel, Spearman ($r_s$ = $-$0.07, \emph{p}-value = 7.67$\times$10$^{-1}$), Pearson ($r_p$ = 0.09, \emph{p}-value = 7.16$\times$10$^{-1}$), and~Kendall (\mbox{$r_k$ = 0.02}, \emph{p}-value = 9.22$\times$10$^{-1}$) for the second panel.

\unskip







\end{document}